\documentclass[11pt]{JHEP3}

\usepackage{latexsym}
\usepackage{amsfonts}
\usepackage{amssymb}
\usepackage{graphics}
\usepackage{graphicx}
\usepackage{epsf}
\def\){\right)}
\def\({\left( }
\def\]{\right] }
\def\[{\left[ }
\def\half{\frac{1}{2}}

\def\bt{\partial_{\bar t}}
\def\t{\partial_t}
\def\no{\nonumber \\}
\def\d{\partial}
\def\db{{\bar \partial}}

\def\be{\begin{equation}}
\def\ee{\end{equation}}
\def\ba{\begin{eqnarray}}
\def\ea{\end{eqnarray}}
\def\no{\nonumber \\}

\def\Z{{\mathbb{Z}}}
\def\C{{\mathbb{C}}}

\def\ra{\rangle}
\def\la{\langle}

\title{Toda systems in closed string tachyon condensation}

\author {Sunggeun Lee and Sang-Jin Sin \\
{\em Department of Physics, Hanyang University, 133-791, Seoul}
}

\abstract{We consider $tt^*$ equations appearing in the study of localized tachyon condensations.  They are 
described by various Toda system when we consider the condensation by the lowest tachyon corresponding to the monomial $xy$. The tachyon potential is calculated as a solution to these equations. The Toda system appearing in the deformation of
$\C^2/\Z_n$  by $xy$ is identical to that of $D_n$ singularity deformed by $x$. 
For $\C^3/\Z_n$ with $xyz$ deformation, we find only generic non-simple form, similar to  the case appearing in $\C/\Z_5\to \C/\Z_3$ and we discuss the difficulties in these cases.
}

\newpage
\begin{document}


\section{Introduction}

\setcounter{equation}{0}

The study of localized tachyon condensation
\cite{aps,vafa,hkmm,dv,many} has been considered with  many
interesting developments. The basic picture is that tachyon
condensation induces cascade of decays of the orbifolds to  less
singular ones until the spacetime supersymmetry is restored.
Therefore the localized tachyon condensation has a geometric
description as the resolution of the  spacetime singularities.

Following the line of Vafa's reformulation of the problem in terms
of Mirror  Landau-Ginzburg theory,  we worked out the detailed
analysis on the fate of spectrum and the background geometry under
the tachyon condensation as well as the question of what is the
analogue of c-theorem with the GSO-projection in a series of
papers\cite{sin,sinlee,gso}.

The study of tachyon dynamics would be greatly improved if we know
the potential that governs the condensation process. In this
direction, Dabholkar and Vafa\cite{dv} proposed that the tachyon
potential is given by the maximal charge and worked out the case
describing the decay of $\C^1/\Z_3$ to $\C$. In the previous paper
\cite{ls}, we generalized the result to $\C^2/\Z_n$  for
 $n=3,4,5$. We encountered solvable non-linear equations called
Painleve III as well as its degenerate case. So it is interesting
to ask which integrable system is waiting us for general $n$'s, if
they are still integrable at all.

In this paper, we consider $\C^2/\Z_n$ as well as to 
$\C^1/\Z_n$ for any $n$ and show that the resulting systems
are various Toda systems when we consider the
condensation by the lowest tachyon corresponding to the monomial
$xy$. We find that the potential is calculated as a solution to
Toda equations. One interesting notice is that the Toda system
appearing as $tt^*$ equation in the deformation of $\C^2/Z_n$  by
$xy$ is identical to that of $D_n$ singularity by $x$, whose
mirror dual geometry is not clear.

The outline of the paper is as follows. In section 2, we summarize the basic concepts to introduce languages. In section 3, we consider the $tt^*$ equations for $\C^1/\Z_n$. In section 4, we calculate the $tt^*$ equations for $\C^2/\Z_n$. In section 5, we discuss and conclude. In appendix A, we calculate for $\C^3/\Z_n \to \C^3$  as well as for $\C\/\Z_5\to \C/\Z_3$ and discuss the difficulties in analyzing the resulting equations quantitatively. In appendix B, we give  a sample residue calculation for topological metric that is an important
ingredient of $tt^*$ equations. 

\section{Landau-Ginzburg description  of orbifold geometry and $tt^*$ equation}
\setcounter{equation}{0}

The Mirror of $\C^2/\Z_n$ is an orbifolded Landau-Ginzburg (LG)
model with potential $ W=x^n+y^n$\cite{hori,wittenN2}. The main
steps for the tachyon potential is to calculate the $tt^*$
equation and their solutions. In this section we summarize the
basic concepts on these material following  \cite{vafa} and
\cite{cv}.

{\it Landau-Ginzburg description of $\C^2/\Z_n$:}
Here we give  a brief summary of Landau-Ginzburg formulation of  localized
tachyon condensation\cite{vafa}. For simplicity we take $\C^2/\Z_n$.
The orbifold $\C^2/\Z_n$ is defined by the  $\Z_n$ action given by
equivalence  relation \be
(X_1,X_2)\sim(\omega^{k_1}X_1,\omega^{k_2}X_2),\quad
\omega=e^{2\pi i /n} .\label{znaction}\ee We call $(k_1, k_2)$ as
the generator of the $\Z_n$ action. The orbifold can be imbedded
into the gauged linear sigma model(GLSM) \cite{wittenN2}. The
vacuum manifold of the latter  is described by the D-term
constraints \be -n|X_0|^2+  k_1 |X_1|^2 +k_2 |X_2|^2=t .\ee Its
$t\to -\infty$ limit corresponds to the orbifold and the
$t\to\infty$ limit is the $O(-n)$ bundle over the weighted
projected space $WP_{k_1,k_2}$. In the latter case, $X_0$
direction corresponds to the non-compact fiber of this bundle and
$t$ plays role of size of the $WP_{k_1,k_2}$.

By dualizing  this GLSM, we get a LG model with  a superpotential\cite{hori}
\be W=  \exp(-Y_1)+\exp(-Y_2)+\exp(-Y_0), \ee
where twisted chiral fields
$Y_i$ are periodic $Y_i\sim Y_i+2\pi i$ and  related to $X_i$ by
$Re[Y_i]=|X_i|^2.$ Introducing  the variable $ u_i:=e^{-Y_i/n},$
the periodicity of $Y_i$ imposes the identification : $u_i \sim
e^{2\pi i/n} u_i $ which necessitate  modding out each $u_i$ by
$\Z_n$. The D-term  constraint is expressed as
$e^{-Y_0}=e^{t/n}u_1^{k_1}u_2^{k_2} $ whose invariance requires
that only one $\Z_n$ can be independent. The result  is usually
described by \be [W= u_1^{n}+u_2^{n}+e^{t/n}  u_1^{k_1}u_2^{k_2}
]// \Z_n.\label{orLGeq} \ee which describe  the  mirror
Landau-Ginzburg model of the linear sigma model. As  $t\to
-\infty$ limit, mirror of the orbifold is \be[W=
u_1^{n}+u_2^{n}]// \Z_n.\ee

>From now on we use $x,y$ instead of $u_1,u_2$ for notational
convenience and fix $(k_1,k_2)=(1,k)$ since it is
allowed\cite{sinlee}.  Since it is  not ordinary Landau-Ginzburg
theory but an orbifolded  version, the chiral ring structure of
the theory is  slightly different from that of LG model. For
example, the   chiral ring of LG model is \be \{x^iy^j |
i,j=1,2,\cdots, n-1\}, \ee
 while that of orbifolded LG model is
\be \{x^iy^j| (i,j)=l(1,k), l=1,2,\cdots, n-1\}, \ee where $(1,k)$
is the generator of the chiral ring and $k$ defines  the orbifold
action in $\C^2$: \be X_1\to \omega X_1, ~~X_2\to \omega^k X_2.\ee
The former has $(n-1)^2$ elements while the latter has the $n-1$
elements. We confine ourselves to the case where we deform the
theory by the lowest charge $xy$.

{\it $tt^*$ equations:} Here we briefly summarize the content of
$tt^*$ equation following \cite{cv}. In an $N=2$ SUSY theory,
there are two supersymmetry charges, ${\bar Q}$ and $Q$. Their
property is \be ({\bar Q})^2=(Q)^2=0,~~~\{{\bar Q},
Q\}=H,~~~({\bar Q})^\dagger= Q, \ee where $H$ is the hamiltonian.
Topological theory is obtained by declaring ${\bar Q}$ to be a
BRST operator and by identifying the BRST cohomology of ${\bar Q}$
with physical Hilbert space. The ground states $|i \ra$ satisfy $
{\bar Q}|i\ra=0,$ where we take the space to be a circle with
periodic boundary condition. The topological operators $\phi_i$
are defined to be operators which commute with ${\bar Q}$, i.e.
$[{\bar Q},\phi_i]=0 ,$ and are called chiral fields. These chiral
fields form a ring because of OPE of two of them is ${\bar
Q}$-closed. \be \phi_i\phi_j=C^k_{ij}\phi_k+ [{\bar Q},\Lambda].
\ee The CPT conjugate operators ${\bar \phi_i}$ commute with $ Q$
and are called anti-chiral. There is an one-to-one correspondence
between the ground states and the chiral operators:
$\phi_i|0\ra=|i\ra + {\bar Q}|\chi\ra$. By definitions, the
resulting state is a topological state. By acting the anti-chiral
operators we get anti-topological state. The topological metric is
defined by the two point function of chiral operators:$
\eta_{ij}=\la j | i\ra, $ and the ground state metric is defined
as an inner product of a topological and an anti-topological
state: $ \la {\bar j} | i\ra =g_{i {\bar j} }$. They satisfy the
reality condition: \be \eta^{-1}g(\eta^{-1})^* = I. \ee The two
point function $\eta_{ij}$ can be calculated by the Grothendieck
residue \ba &&\eta_{ij}=\la \phi_i \phi_j \ra =Res[\phi_i \phi_j],
\no &&Res[\phi]={ 1 \over {(2\pi i)^n}} \int_\Gamma {
{\phi(X)dX^1\wedge \cdots \wedge dX^n } \over {
\partial_1W
\partial_2W \cdots \partial_nW } }, \ea where $W$ is the
superpotential.

Now let us consider the perturbation of the action
\be
S=\int d^2 z d^4\theta K +( \int d^2 z d^2 \theta W + c.c.)
\ee
by the chiral operators with
\be
\delta S =\delta t_i \int d^2z d^2 \theta  \phi_i + c.c.,
\ee
where $t_i$ correspond to the (complex) couplings in the
theory.
As $t_i$ change the Ramond vacua change.
The method of the computation of the metric $g_{i {\bar j}}$ is to
study its behavior under perturbatuions which preserves the
$N=2$ SUSY.

Analogous to Berry connection let us introduce the connections $A_i$
on the coupling constant space
(holomorphic parameter space)  given by
\be
A_{ia{\bar b}}=\la {\bar b}|\partial_i |a\ra .
\ee
This connection measures the way the ground state subsector
varies in the Hilbert space as the couplings change.
Under a coupling constant dependent change of
basis for the vacua, $A_i$ transforms as a gauge field.

If we  define the covariant derivative by $D_i=\partial_i -A_i$,
then the metric $g_{i{\bar j}}$ is covariantly constant $ D_i
g_{i{\bar j}}=0={\bar D}_ig_{i{\bar j}}. $ One can prove the
following equations\cite{cv} \ba &&[D_i,D_j]=[{\bar D}_i,{\bar
D}_j]=0, \no &&[D_i,{\bar D}_j]=-[C_i,{\bar C}_j], \ea which are
called $tt^*$ equations. If we choose holomorphic gauge $A_{\bar
i}=0$ the second equation can be written as \be {\bar
\partial}_i(g\partial_j g^{-1})-[C_j,g(C_i)^\dagger g^{-1}]=0, \ee
which will be our main concern here.


\section{$tt^*$ equations for $\C^1/\Z_n$}

\setcounter{equation}{0}

The $tt^*$ equations for the one field LG
model under the deformation of most relevant operators ($x,x^2$)
were calculated by Cecotti and Vafa\cite{cv} and the results were
various Toda systems. One can adopt their result to discuss the
localized tachyon condensation in $\C^1/\Z_n$. The only thing one
has to be careful is the renaming $n+1\to n$ and the shift in power in the  chiral ring elements associated with shift $\d_x\to x\d_x$. As a consequence, the evenness 
and oddness of various quantities are all reversed. So it is
useful to rewrite the $tt^*$ equations in the present context and present notations.

\subsection{Deformation by $x$: condensation of the most tachyonic state}

First we discuss the tachyon condensation by the most tachyonic
state which has mass $m^2=2\alpha'(1/n-1)$. The monomial
representation of the corresponding  operator is simply given by
$x$ whose NS-charge is $1/n$. So the tachyon condensation
in the LG description is described  by the superpotenial \be
W={x^n \over n} -tx. \ee Since the basic variable is $\log x$, the
chiral ring is \be {\cal R}=\C[x]/\d_{\log x} W=
\{x,x^2,\cdots,x^{n-1}\}. \ee The non-vanishing components of the
topological metric $\eta$ are \be \eta_{i,n-i}=1, ~for ~
i=1,2,\cdots,n-1. \ee Let $a_i:=g_{i{\bar i}}=\la {\bar i}|i \ra$.
The reality condition gives \ba
 a_ia_{n-i}&=& 1 ,\no
if ~ n=2m+1, ~ a_m &=&1 \ea The non-vanishing components of
structure constant matrix $C_t$ are \be
(C_t)_{1}^2=(C_t)_{2}^3=\cdots=(C_t)_{n-2}^{n-1}=1,
~(C_t)_{n-1}^{1}=t. \ee With these, the $tt^*$ equations can be
calculated. \footnote{For $\C^1/\Z_n$ problem, some of the $tt^*$
equations found by  Cecotti and Vafa in \cite{cv} can be adopted
for our purpose, while for $\C^2/\Z_n$ we really have to workout
the corresponding systems.} \ba &&-\partial_{\bar t}\partial_t
\log a_1 ={a_2 \over a_1} - |t|^2 {a_1 \over a_{n-1}} \no
&&-\partial_{\bar t}\partial_t \log a_i ={a_{i+1}\over a_i}-{a_i
\over a_{i-1}},~~~i=2,\cdots,n-2\no &&-\partial_{\bar t}\partial_t
\log a_{n-1} = |t|^2{a_1 \over a_{n-1}} -{a_{n-1} \over a_{n-2}}
\label{tta}. \ea By introducing new variables
 \ba q_i&=&\log
a_i-\frac{2i-n}{2(n-1)}\log |t|^2,\no
 z&=& \frac{n-1}{n} t^{n/(n-1)},
 \ea
 and $q_n=q_1, ~q_0=q_{n-1}$, the equations given in
 eqs.(\ref{tta}) take the form \be \d_z\d_{\bar
z}q_i+e^{q_{i+1}-q_i}-e^{q_i-q_{i-1}}=0, ~for~ i=1,\cdots, n-1.
\ee which is the $A_{n-2}$ Toda equations. By imposing the reality
condition $q_i+q_{n-i}=0$, we can reduce the system to $C_m$ (if
$n=2m+1$) or $BC_m$ (if $n=2m+2$) Toda system.

\subsection{Deformation by $x^2$}

\be
W={x^n \over n} -{t\over2}x^2.
\ee
The chiral ring is the same as before.
The   topological metric $\eta$ are
\be
\eta_{i,j}=\delta_{i+j,n}+t\delta_{i,n-1}\delta_{j,n-1}, ~for ~ i,j=1,2,\cdots,n-1.
\ee
The non-vanishing components of the ground state metric $g_{ij}$ is
\be
g_{i,{\bar i}}=a_i, ~~g_{n-1,{\bar 1}}=b,~~ g_{1, \overline{n-1}}={\bar b}.
\ee
The reality condition gives
\ba
 a_ia_{n-i}&=& 1 ,~for ~ i=2,\cdots,n-2  \no
a_{n-1}&=&1/a_1+{|t|^2 \over 4}a_1 ,\no b&=& \half t a_1, \no and
 ~ ~~~a_m &=&1 ~if ~n=2m+1, \ea The non-vanishing components of structure
constant matrix $C_t$ are \be
(C_t)_{1}^3=(C_t)_{2}^4=\cdots=(C_t)_{n-3}^{n-1}=1,
~(C_t)_{n-2}^{2}=(C_t)_{n-1}^{3}=t. \ee According to whether $n$
is even or odd, the $tt^*$ equations are of different type.

For $n=2m+1$, we define
\ba
q_i&=&-\log a_{2i-1} ~~for ~~i=1,2,\cdots,[(m+1)/2],\no
   &=& \log a_{2(m-i)+2} ~~for ~~i=[(m+1)/2]+1,\cdots,m
\ea
\def\dbar{{\bar \partial}}
Then the equations are \ba \d\dbar q_1 &=&
e^{q_1-q_2}-{1\over4}|t|^2e^{-(q_1+q_2)},\no \d\dbar q_2 &=&
e^{q_2-q_3}-e^{q_1-q_2}-{1\over4}|t|^2e^{-(q_1+q_2)},\no \d\dbar
q_i &=& e^{q_i-q_{i+1}}-e^{q_{i-1}-q_i} ,\no \d\dbar q_m &=&
|t|^2e^{2q_{m}}-e^{q_{m-1}-q_m}. \ea It is easy to re-scale the
variables to make the equation as  a standard ${\tilde
B}_m:=D^T(SO(2m+1))$ Toda form.

For even $n$, the $tt^*$ equations decouples into two  independent
Toda systems and we have  two cases according to  whether $n/2$ is
even or odd.
\begin{enumerate}
    \item $n=4m$: $a_1,a_3,\cdots,a_{2m-1}$ satisfies
    ${\tilde B}_m$ Toda equations, while $a_2,a_4, \cdots,a_{2m-2}$
    satisfies $BC_{m-1}$ Toda.
    \item $n=4m+2$: $a_1,a_3,\cdots,a_{2m-1}$ satisfies $B_m$ Toda
    equations, while $a_2,a_4, \cdots,a_{2m}$ satisfies $C_m$ Toda.

\end{enumerate}


\section{$tt^*$ equations for $\C^2/\Z_n$}

\setcounter{equation}{0}
Now we consider $\C^2/\Z_{n(1,1)}$ with
the generator $xy$ and
 we consider the condensation of most tachyonic operator,
 namely the generator $xy$ itself. Then the superpotential is given by
\be W={x^n \over n} + {y^n \over n} -   t xy. \ee Since the $tt^*$
equations for this case has never been calculated, we do it here.
Before we consider the mirror of $\C^2/\Z_n$, namely the
orbifolded LG model, we first consider the generality of LG model
itself without the orbifold action.

\subsection{General aspects of LG model without orbifold action}

To see what is the non-vanishing elements of the ground state
metric $g_{i\bar j}$ and the topological metric $\eta_{ij}$, we
first consider the discrete symmetries of the superpotential.

Under the transformation \be x\to \omega^a x ~{\rm and}~
y\to\omega^b y \ee the superpotential $W$ is symmetric upto an
over all phase $\omega^{*}$ which is cancelled by transforming the
$\theta \to \omega^{(a+b)/2}\theta$. By requiring invariance of
the action we have \be \omega^{an}=\omega^{bn}=\omega^{a+b}. \ee
Without loss of generality, we can set: \be a=1, ~~~b=n-1,
~~~\omega^{n(n-2)}=1, \ee up to an equivalence class. Now
$\eta_{ij}$  is given by the Griffith residue \be \eta_{ij}= \la
x^{i_1}y^{j_1}\cdot x^{{i}_2}y^{{j}_2}\ra =\int {d x \over
x}\wedge {dy \over  y}{ x^{i_1+i_2}y^{j_1+j_2}  \over
{(x^{n}-txy)(y^{n}-txy)}}. \ee The non-vanishing components should
be invariant under the above symmetry transformations. This
requires \be (i_1+i_2-2)+(n-1)(j_1+j_2-2)\equiv 0~ {\rm mod}
~n(n-2). \ee The selection rule is given by the solutions to this
equation: \be i_1+i_2=j_1+j_2=I \equiv 2~ {\rm mod} ~(n-2). \ee
For the actual value, we perform explicit residue calculations,
whose example is given in the appendix. We found  that
\begin{itemize}
    \item  $I=2$ case:  We have one solution $i_1=i_2=1, j_1=j_2=1.$
    However, it turns out that $\eta=0$  in this case.
    \item   $I=n$ case: $i_1=j_1=i, i_2=j_2=n-1-i, i=1,\cdots,n-1.$
These are $n-1$ cross diagonal elements.  All of them are 1.
    \item  $I=2(n-1)$ case: There is only one element $i_1=j_1=i_2=j_2=n-1,$
     and the value is $\eta_{ij}=t^2$.
\end{itemize}

One can find similar selection rule for the metric element
$g_{i\bar{j}}= \la x^{i_1}y^{j_1}x^{\bar{i}_2}y^{\bar{j}_2}\ra$.
For non-vanishing elements we need \be
\omega^{i_1-i_2+(n-1)j_1-(n-1)j_2}=1 ~{\rm with
}~~\omega^{n(n-2)}=1, \ee which gives \be
i_1-i_2+(n-1)(j_1-j_2)\equiv 0~{\rm mod}~n(n-2). \ee As
consequences we have \be i_1-i_2=j_1-j_2~~ \equiv 0~~ {\rm mod}
~(n-2).\ee
 There are two cases:
\begin{itemize}
    \item $i_1=i_2,~  j_1=j_2$. These are the diagonal elements of $g_{i{\bar j}}$.
We define
$a_{ij}:=\la {\bar x}^{i}{\bar y}^{j}|x^{i}y^{j}\ra$.
    \item $j_1-j_2=i_1-i_2=\pm (n-2)$, namely
\ba
i_1=j_1=n-1, && ~ i_2=j_2=1,  ~or \\
 i_1=j_1=1,&&  ~ i_2=j_2=n-1, \ea
which are the two corner-most off-diagonal elements.
we call it as $\bar b$ and $b$ respectively.
\end{itemize}
The calculation of $tt^*$ equation for general LG model  is
cumbersome. We now simplify our life by performing the orbifold
action. The simplification is given by reducing the number of
chiral ring elements from $(n-1)^2$ to $n-1$.

\subsection{Orbifolded LG: the mirror of $\C^2/\Z_n$}

The chiral ring in this case is generated by $xy$:
\be
{\cal R}=\{xy, (xy)^2,\cdots, (xy)^{n-1} \}.
\ee
Let \be
a_i=\la {\bar x}{\bar y})^i|(xy)^i\ra=e^{-q_i }.\ee

The topological metric can be calculated to give
\be
\eta=\eta_{ij}=\delta_{i+j, n}+t^2\delta_{i,n-1}\delta_{j,n-1}.
\ee
The  structure constants are
\be
(C_t)_i^j=\delta_{i,j-1}+t^2\delta_{i,n-1}\delta_{j,2}.
\ee
The reality condition gives
\be
a_ia_{n-i}=1~ (for ~i=2,\cdots,n-1), ~~~~~a_{n-1}=1/a_1+|t|^4a_1/4.
\ee
With all these, we can calculate the $tt^*$ equations.
Let's introduce a parameter $s$ by
\be
n=2m+2-s, ~~ s=0,1.\ee
Then $tt^*$ equation can be written as
\ba
&&-\partial_{\bar t}\partial_t \log a_1 ={a_2 \over a_1} -{1\over 4}|t|^4
a_1 a_2 \no
&&-\partial_{\bar t}\partial_t \log a_2 ={a_3 \over a_2}-{a_2 \over a_1}-{1\over 4}|t|^4
a_1 a_2 \no
&&-\partial_{\bar t}\partial_t \log a_i ={a_{i+1}\over a_i}-{a_i \over a_{i-1}},~~~i=3,\cdots,m-1\no
&&-\partial_{\bar t}\partial_t \log a_m =a_m^{-(1+s)} -{a_m \over a_{m-1}}
\ea

Let's re-scale the variables to eliminate $|t|$'s and other
coefficients. \be a_i =\delta_i
|t|^{\alpha_i}e^{-q_i},~~\zeta=\gamma t^\beta. \ee Then \ba
&&\alpha_1 =-2,~~\alpha_2=2\beta-4  \no &&\alpha_m =(m-1)(2\beta
-2)+\alpha_1 =-{2(\beta -1) \over {1+s}}, \no &&\beta=1+ { {1+s}
\over  {1+(1+s)(m-1)}} \no &&\alpha_k =-2{ {(1+s)(m-k)+1}\over
{1+(1+s)(m-1)}} \no &&\delta_1=2,~~\delta_m=\delta_1 (\gamma
\beta)^{2(m-1)}=(\gamma \beta)^{-{1\over {1+s} }} \no &&
\gamma=\beta^{-1}\({1\over 2}\)^{ {{1+s} \over {1+2(m-1)(1+s)} }}
\ea Then \ba \d \db q_1 &=& e^{q_1-q_2}-e^{-(q_1+q_2)},\no \d \db
q_2 &=& e^{q_2-q_3}-e^{q_1-q_2}-e^{-(q_1+q_2)},\no \d \db q_i &=&
e^{q_i-q_{i+1}}-e^{-(q_{i-1}-q_i)}, ~i=3,\cdots,m-1 \no \d \db q_m
&=& e^{(1+s)q_m}- e^{(q_{m-1}-q_m)} \ea
One can bring this
equation to the Toda equations: for $s=0$ (even $n$) these are
$B_m$ Toda and for $s=1$ (odd $n$) they are $\tilde
B_m=D^T(SO(2m+1))$ Toda system. It is a curious fact that the Toda
system  appearing in the deformation of $\C^2/Z_n$  by $xy$ is
identical to that appearing in the deformation of $D_n$
singularity ($W={1\over n-1}x^{n-1}+xy^2$) by $x$. Notice that the
mirror geometric correspondence of $D_n$ singularity  has not been
clear at all and is still not clear.

The charge matrix $Q=g\d_\tau g^{-1}-1$ with $\tau=\log\lambda$
can be calculated to be given by
\be  Q=\left(%
\begin{array}{ccccc}
   \frac{2-n}{n}+a_{11}\d_\tau a_{11}^{-1}  & 0&\dots & 0 & 0 \\
  0&  \frac{4-n}{n}+a_{22}\d_\tau a_{22}^{-1}  & \dots &0 & 0 \\
    \dots&   \dots&   \dots & \dots &  \dots \\
  0 & 0 &  \dots& \frac{n-4}{n}-a_{22}\d_\tau a_{22}^{-1} &0  \\
  t^2 a_{11}\d_\tau a_{11}^{-1}  & 0 &  \dots &0 & \frac{n-2}{n}-a_{11}\d_\tau a_{11}^{-1} \\
\end{array}%
\right).
\ee
 Notice that we do not need to perform the diagonalization to get the eigenvalues of $Q$.
 In terms of $q_{ij}$ and $\lambda(=\zeta)$, and if we look at the $|\lambda|=z$
 dependence only, the tachyon
potential can be identified as  \be V=2Q_{max}= -
z\d_{z}q_{11}(z). \ee So far, no mathematical literature on the
solution to the equation is available.  However, from the physical
intuition that in the final stage of tachyon condensation there is
no nontrivial chiral primaries with charge other than 0 and also
from the experience from the low $n$ cases, we expect that the
potential monotonically decrease from the value $2/n-1$ at $t=0$
to $0$ at $t\to \infty$.

\section{Discussion}

In this paper, we worked out equations describing the tachyon
condensation in orbifolds $\C^1/\Z_n$ and $\C^2/\Z_n$ and the
tachyon potential. The resulting equations can be identified as
various Toda systems when we consider the condensation by the
lowest tachyon. We find that the potential is calculated as a
solution to Toda equations and conjectured to be monotonically
decreasing.

 It is interesting to notice that the Toda system
appearing as $tt^*$ equation in the deformation of $\C^2/\Z_n$ by
$xy$ is identical to that of $D_n$ singularity by $x$. It is not
clear whether this signalize the possibility to identify the
geometry corresponding to $D_n $ singularity as $\C^2/\Z_n$. It is
tantalizing problem to answer to the question what is the
geometry corresponding to the general singularities along the line
of the idea that $A_{N-1}$ singularity corresponds to the
$\C/\Z_{N}$.

We should point out that we considered the string theories without
GSO projection only when we discuss the $xy$ deformation of
$\C^2/\Z_n$ and $x^2$ deformation of $\C^1/\Z_n$. The reason is
simply that if we impose the type II projection rule, the above
operators are  projected out and remaining spectrum gives us
complicated reality conditions so that the resulting $tt^*$
equations are technically beyond our reach so far. Related problem
arises when the daughter theory is also an orbifold $\C^1/\Z_n$
background.

 One of the prime difficulty in the solution comes from the constraints coming from the reality condition. For the lowest deformations of $\C/Z_n$ and $\C^2/\Z_n$, the reality conditions can be easily solved as we have seen in the main sections.
However, for $\C^3/\Z_n$ with lowest perturbation $xyz$ or for $\C/\Z_n$ case with non-trivial deformation like $x^3$, the solutions are non-trivial. If we adopt an special ansatz, it the resulting solutions corresponds to the wrong boundary conditions
as we described at the end of  the appendices.

Of course most immediate question is whether we can write down
analytic solution of the toda systems we found with appropriate
boundary conditions. So far no mathematical literature is
available to our knowledge. We wish to report on these issues in
later publications.

\setcounter{equation}{0}
\appendix
\section{$\C^3/\Z_n$}
In this appendix we present a somewhat incomplete result on $\C^3/\Z_n$. We spent long time without success in getting a simple system with solutions satisfying proper boundary conditions. We point out what are the difficulty through an example $n=5$ case. We also give a comparison with similar system
 $\C/\Z_5\to \C/\Z_3$. 

The superpotential is 
\be
W=x^n/n+y^n/n+z^n/n-txyz.
\ee
We work in the basis
\be
\{  xyz, (xyz)^2,\cdots,(xyz)^{n-1} \}.
\ee
Under the transformation
\be
x\to \omega^a x,~~~y\to \omega^b y,~~~z\to \omega^c z,
\ee
with well chosen $a$, $b$, and $c$, the superpotential
becomes invariant. The choice is
\be
b=a, ~~~c=a(n-2), ~~~~with~\omega^{an(n-3)}=1. \label{omega}
\ee

From this we can get the non-vanishing components of
the topological metric $\eta_{ij}$ and metric $g_{{\bar j}i}$.
First for $\eta_{ij}$,
\be
\eta=\int { {(xyz)^I dx\wedge dy\wedge dz} \over {
(x^{n-1}-tyz)(y^{n-1}-txz)(z^{n-1}-txy)} }.
\ee
In order for this to be invariant under eq.(\ref{omega}),
\be
an(I-1)=0~{\rm mod} ~an(n-3)
\ee
which means some components of $\eta$ are non-zero.
One can find and calculate the non-vanishing values of $\eta$:
\be
0:~~when~I=1~and~n-1:,~~~~~t^3:~~when~I=2n-3.
\ee

For the metric
\be
g_{{\bar j}i}=\la (xyz)^{\bar j}|(xyz)^i \ra
\ee
to be invaraint we obtain
\be
an(i-j)=0,~{\rm mod}~ an(n-3).
\ee
Therefore we obtain the diagonal of the metric (we spell out them as
$a_{ii}$) and the off-diagonal components (denoted by $b_1$ and $b_2$ and
their conjugates).

One can find $C_t$:
\be
(C_t)^2_1=(C_t)^3_2=\cdots=(C_t)^{n-2}_{n-3}=1,~~~(C_t)^{n-1}_3=t^3.
\ee

The difficult part is the reality condition.
Although we can solve it with the ansatz
\be
b_1=a_{11} t^3/2,~~~~b_2=a_{22}t^3/2,
\ee
there is no guarantee that its consequence is consistent with the boundary conditions we have to impose. 
With all these we can write down the $tt^*$ equation, but for $n \geq 6$
we get inconsistent equation like
\be
a_{11}=0~(n=6)~~~{\rm or}~~a_{11}a_{33}=0~(n>6)
\ee
which violates regularity of the metric.
The  $tt^*$ equation for general $\C^3/\Z_n$ is not simple. 
Main reason is the absence of the simple solution to the reality constraint. We give some detail for $n=5$.

\subsection{ $tt^*$ euqation for $\C^3/\Z_5 \to \C^3/\Z_3$}

In this case the superpotential is written as
\be
W=x^5/5 + y^5/5 +z^5/5 - t xyz.
\ee
The symmetry that the superpotential has is
\be
x\to \omega^a x~~y\to \omega^a y ~~z\to
\omega^{3a}z~~~with~~\omega^{10a}=1.
\ee
This restricts the topological metric and metric to be
\be
\eta_{ij}=\pmatrix
{0 & 0&0&1 \cr
0&0&1&0 \cr
0&1&0&t^3 \cr
1&0&t^3&0
},
~~~~~g_{{\bar j}i}=\pmatrix
{a_{11}&0&\bar{b}_1&0\cr
0&a_{22}&0&\bar{b}_2 \cr
b_1&0&a_{33}&0\cr
0&b_2&0&a_{44}
},
\ee
and the multiplication matrix for the perturbing chiral ring element $xyz$ is
\be
(C_t)^j_i=\pmatrix
{0&1&0&0\cr
0&0&1&0\cr
0&0&0&1\cr
0&0&t^3&0
}.
\ee
We take the basis as
\be
\{ (xyz),(xyz)^2,(xyz)^3,(xyz)^4\}.
\ee

We list here all $tt^*$ equations for $\C^3/\Z_5$ case.
\ba
&&\partial_{\bar t}\( { {a_3 \partial_t a_1 -b_1 \partial_t {\bar b}_1}\over {-a_1 a_3 +b_1{\bar b}_1} } \)=
{ {a_2 a_3 -b_1 {\bar b}_2} \over {a_1a_3-b_1 {\bar b}_1} },~~~~
\partial_{\bar t}\( { {b_1 \partial_t a_3 -a_3 \partial_t b_1} \over
{a_1 a_3-b_1{\bar b}_1 }}\) ={ {a_4 b_1 -a_3 b_2}\over {-a_1 a_3 +b_1 {\bar b}_1} },\no
&&\partial_{\bar t} \( { {a_4 \partial_t a_2 -b_2 \partial_t {\bar b}_2 } \over
{-a_2 a_4 +b_2 {\bar b}_2 } }\)=\no
&&~~~~~{ {-a_2^2 a_3 a_4 +a_1 a_3^2(a_4-{\bar
t}^3b_2)+a_2(a_4b_1+a_3b_2){\bar b}_2-b_1(a_3{\bar b}_1(a_4-{\bar t}^3b_2)+b_2{\bar b}_2^2) }
\over {(a_1a_3-b_1{\bar b}_1)(a_2a_4-b_2{\bar b}_2)} },\no
&&\partial_{\bar t}\( { {b_2 \partial_t a_4 -a_4 \partial_t b_2}\over {a_2 a_4 -b_2{\bar b}_2} } \)=\no
&&~~~~{ {-t^3 a_3 a_4 b_1{\bar b}_1+|t|^6a_3 b_1 {\bar b}_1b_2+t^3 a_1
a_3^2(a_4 -{\bar t}^2b_2)+a_2a_4(a_4b_1-a_3b_2)-a_4b_1b_2{\bar b}_2+a_3b_2^2{\bar b}_2}
\over {(a_1 a_3-b_1{\bar b}_1)(a_2a_4-b_2{\bar b}_2)} },\no
&&\partial_{\bar t}\( { {{\bar b}_1 \partial_t a_1 -a_1 \partial_t {\bar b}_1}\over {a_1 a_3 -b_1{\bar b}_1} } \)=\no
&&~~~~~~~~~~~~~{ {a_1(-b_2{\bar b}_2^2+a_3{\bar b}_1(-a_4+{\bar t}^3{\bar
b}_2)+a_2(-|t|^6a_3{\bar b}_1+a_4{\bar b}_2)}\over {(a_1a_3-b_1{\bar
b}_1)(a_2a_4-b_2{\bar b}_2)} }  \no
&&~~~~~~~+ { {{\bar b}_1(-a_2^2a_4+b_1{\bar b}_1(a_4-{\bar
t}^3b_2-t^3b_2)+a_2(|t|^6b_1{\bar b}_1+b_2{\bar b}_2))}
\over {(a_1a_3-b_1{\bar b}_1)(a_2a_4-b_2{\bar b_2}) } },\no
&&\partial_{\bar t}\( { {a_1 \partial_t a_3 -{\bar b}_1 \partial_t  b_1}\over {-a_1 a_3 +b_1{\bar b}_1} } \)=\no
&&~~~~~~~~~{ {{\bar b}_1(a_2(|t|^6a_3b_1-a_4b_2)+b_2^2{\bar b}_2+a_3b_1(a_4-{\bar
t}^3b_2-t^3{\bar b}_2))}\over { (a_1a_3-b_1{\bar b}_1)(a_2a_4-b_2{\bar
b}_2) }}\no
&&~~~~~+ {{a_1(a_2(-|t|^6a_3^2+a_4^2)-a_4b_2{\bar b}_2+a_3^2(-a_4+{\bar
t}^3b_2 +t^3{\bar b}_2))}
\over { (a_1a_3-b_1{\bar b}_1)(a_2a_4-b_2{\bar b}_2) } },\no
&&\partial_{\bar t}\( { {{\bar b}_2 \partial_t a_2 -a_2 \partial_t {\bar b}_2}\over {-a_2 a_4 +b_2{\bar b}_2} } \)=\no
&&~~~{ {a_1(a_2({\bar t}^3a_3^2-a_4{\bar b}_2)+{\bar b}_2(-a_3^2+b_2{\bar
b}_2))+{\bar b}_1(a_2^2a_4+a_3b_1{\bar b}_2-a_2({\bar t}^3a_3b_1+b_2{\bar b}_2))}
\over { (a_1a_3-b_1{\bar b}_1)(a_2a_4-b2{\bar b}_2)} },\no
&&\partial_{\bar t}\( { {a_2 \partial_t a_4 -{\bar b}_2 \partial_t b_2}\over {-a_2 a_4 +b_2{\bar b}_2} } \)=\no
&&~~~{ { a_1(a_2(|t|^6a_3^2-a_4^2)-(t^3a_3^2+a_4b_2){\bar b}_2+{\bar
b}_1(a_2(-|t|^6a_3b_1+a_4b_2)+(t^3a_3b_1-b_2^2){\bar b}_2) }\over
{(a_1a_3-b_1{\bar b}_1)(a_2a_4-b_2{\bar b}_2)} }.\no
\ea
These equations should be supplemented by the 
constraints coming from the 
reality condition:
\ba
(\bar{t}^2a_1+{\bar b}_1)(-t^3a_2+b_2)+a_1(a_4-t^3{\bar b}_2)&=&1,\no
~~a_2(a_3-t^3{\bar b}_1)+(-t^3a_1 +b_1)(-{\bar t}^3a_2+{\bar b}_2)&=&1, \no
a_2(a_3-{\bar t}^3b_1)+b_1{\bar b}_2&=&1,\no 
{\bar b}_1b_2+a_1(a_4-{\bar t}^3b_2)&=&1, \no
a_2{\bar b}_1+a_1(-{\bar t}^3 a_2 +b_2)&=&0, \no
(a_3-{\bar t}^3b_1)(-t^3a_2 +b_2)+b_1(a_4 -t^3{\bar b}_2)&=&0, \no
(a_3-t^3{\bar b}_1)b_2+(-t^3a_1+b_1)(a_4-{\bar t}^3b_2)&=&0.
\ea

We may wonder whether we can find some  ansatz to make the $tt^*$ equations  simple.  We will try an obvious one and show how it does not work. Let's try an ansatz given by 
\be
b_1=a_1 t^3/2,~~~~b_2=a_2 t^3/2, \label{ansatz1}.
\ee
Together with the reality condition above ansatz  gives
\be
a_3=1/a_2 + |t|^6 a_1/4,~~~~a_4 =1/a_1 + |t|^6 a_2/4,
~~~~{\rm with}~a_1a_4=a_2a_3.
\ee
From $tt^*$ equation we get the relation
\be
a_2=-4/(|t|^6 a_1).
\ee
Then finally we get 
\be
-\partial_{ {\bar t}}\partial_t \log a_1=-4/(|t|^6 a_1^2). \label{ttc3}
\ee
Now let us define 
\be
\zeta = t^{-2} ~~and~~a_1 =e^\phi, 
\ee 
the eq.(\ref{ttc3}) becomes Liouville equation
\be
\partial_{\bar{\zeta}}\partial_\zeta \phi=e^{-2\phi}.
\ee
This has an analytic solution.
However this turns out to have  incorrect behavior.

Let us rescale
\be
x=\lambda^{1/5}\tilde{x},~~y=\lambda^{1/5}\tilde{y},
~~z=\lambda^{1/5}\tilde{z}~~\lambda=t^{2/5}.
\ee
Then we have the super potential  is rescaled to give 
\be
\tilde{W}=\lambda \( \tilde{x}^5/5 +\tilde{y}^5/5+\tilde{z}^5/5
-\tilde{x}\tilde{y}\tilde{z} \).
\ee
The charge can be calculated by the relation
\be
Q=\tilde{g}\partial_\tau \tilde{g}^{-1}-3/2.
\ee 
The correlation functions are rescaled to give 
\be
a_i=\la (\bar{x}\bar{y}\bar{z})^i|(xyz)^i\ra=|\lambda|^{6i/5}
\la(\bar{\tilde{x}}\bar{\tilde{y}}\bar{\tilde{z}})^i|(
\tilde{x}\tilde{y}\tilde{z})^i\ra:=|\lambda|^{6i/5}b_i.
\ee
where $b_i$ are components of $\tilde{g}$.
Therefore the charge is
\be
Q_{min}={1\over 2}b_1 |\lambda|\partial_{|\lambda|}
b_1^{-1}-{3\over 2} 
=-{9\over 10}-{1\over 2}|\lambda|{ da_1 \over {d|\lambda|}}
=-{9\over 10}+{5\over 2}|\zeta| {da_1 \over {d|\zeta|}},
\ee
where we have used $\lambda=t^{2/5}$ and $\zeta=t^{-2}$.
So, the required boundary conditions are 
\be
|\zeta|{ da_1 \over {d|\zeta|}}\to 0 \;\;if \;\; |\zeta|\to\infty \; (UV),\;\;
|\zeta|{  da_1 \over {d|\zeta|}}\to \frac{18}{50} \;\;if\;\; |\zeta|\to 0\;(IR). \ee
The general solution for Liouville equation
\be -\d_\zeta\d_{\bar \zeta} \phi + e^{-2\phi} = \frac{1-a}{2}\delta^{(2)}({z}),
\ee
is \cite{seiberg}
\be
e^{-2\phi}= { \partial A(\zeta)\partial B(\bar {\zeta})
\over {(1-A(\zeta)B(\bar{\zeta}))^2}}.
\ee
There are three types of solutions according to the region of $a$:
Elliptic ($a$ real), Parabolic ($a\to 0$), Hyperbolic ($a$ pure imaginary). However, none of such solutions can match the boundary conditions.
Therefore the ansatz given in eq.(\ref{ansatz1}) does not work.
 
\subsection{$tt^*$ euqation for $\C/\Z_5 \to \C/\Z_3$}

Similar problem arise when we study $\C^1/\Z_5$ with   non-trivial  tachyon perturbation, namely,
\be
W=x^5/5-t x^3/3.
\ee
Therefore the basis is
\be
\{ x,x^2,x^3,x^4\}.
\ee
The symmetry restricts the topological metric and metric
which is given by
\be
\eta=\pmatrix 
{ 0&0&0&1 \cr
  0&0&1&0 \cr
0&1&0&t \cr
1&0&t&0 },
~~~~~g=\pmatrix
{a_1&0&\bar{b}_1&0 \cr
0&a_2 &0&\bar{b}_2 \cr
b_1&0&a_3 &0\cr
0&b_2&0&a_4 },
\ee
and $C_t$ is given by
\be
C_t=\pmatrix
{0&0&0&1 \cr
0&0&t&0\cr
0&0&0&t \cr
0&0&t^2&0
}.
\ee
We take ansatz to solve reality condition
\be
b_1=a_1 t/2,~~~~~b_2=a_2 t/2, \label{ans53}
\ee
and the result is
\be
a_3=1/a_2 + |t|^2 a_1/4,~~~~~a_4=1/a_1+ |t|^2 a_2/4,~~~{\rm with}~a_1a_4=a_2a_3.
\ee
$tt^*$ equation gives one more restriction to $a_2$ 
\be
a_2=|t|a_1. 
\ee 
Then we finally get the simplified $tt^*$ equation 
\be
-\partial_{ {\bar t}}\partial_t \log a_1 =
1/a_1^2 -|t|^6a_1^2/16.
\ee
If we define $a_1=1/y$, then the equation becomes
\be
\partial_{ {\bar t}}\partial_t \log y =
y^2 -|t|^6/{16y^2}.
\ee
By introducing change of variables
\be
\zeta =(2/5)t^{5/2},~~~y=\sqrt{1/2}|t|^{3/2}e^{u/2},
\ee
one can rewite this equation as sinh-Gordon
\be
\partial_{{\bar \zeta}}\partial_\zeta u=\sinh u.
\ee
The solution to this equation is describing the solution  whose associated charges decrease to zero, which is not what we want here.  Hence we see that the ansatz (\ref{ans53}) above is not a proper one. 
Then the equation is  given by 
$tt^*$ equation
\ba
&&\bt \( { {a_3 \t a_1-b_1 \t {\bar b}_1 } \over {-a_1a_3+|b_1|^2 } } \)
={ {a_4(a_3-{\bar t} b_1)}\over {a_1a_3-|b_1|^2} }, \no
&&\bt \( { {b_1\t a_3 -a_3 \t b_1} \over {a_1a_3 -|b_1|^2} } \)
={ {ta_4(a_3-{\bar t} b_1)} \over {a_1a_3-|b_1|^2} },\no
&&\bt \( { {a_4\t a_2 -b_2 \t {\bar b}_2} \over {-a_2a_4+|b_2|^2} } \)
={ {|t|^2a_3(a_4-{\bar t} b_2)} \over {a_2 a_4-|b_2|^2} },\no
&&\bt \( { {b_2 \t a_4 -a_4 \t b_2} \over {a_2 a_4 -|b_2|^2} } \)
={ {t|t|^2 a_3(a_4-{\bar t}b_2) }\over { a_2a_4-|b_2|^2} },\no
&&\bt \( { { {\bar b}_1\t a_1 -a_1\t {\bar b}_1} \over
{a_1a_3-|b_1|^2} }\) ={ {{\bar t}a_1(a_2(a_4^2-|t|^2ta_3{\bar 
b}_1)-a_4|b_2|^2+ta_3{\bar b}_1(-a_4+{\bar t}b_2+t{\bar b}_2))}
\over { (a_1a_3-|b_1|^2)(a_2a_4-|b_2|^2) } }\no
&&~~~~~~~~~~+{ { {\bar b}_1(a_2(-a_4^2+|t|^4|b_1|^2)-|t|^2|b_1|^2({\bar 
t}b_2+t{\bar b}_2+a_4(|t|^2|b_1|^2+|b_2|^2))} \over
{(a_1a_3-|b_1|^2)(a_2a_4-|b_2|^2)} }, \no
&&\bt \( { {a_1\t a_3 -{\bar b}_1\t b_1}\over {-a_1a_3 +|b_1|^2} }\)=
{ {t({\bar t}a_1(a_2(-|t|^2a_3^2+a_4^2)-a_4|b_2|^2+a_3^2(-a_4+{\bar 
t}b_2+t{\bar b}_2)))}\over {(a_1a_3-|b_1|^2)(a_2a_4-|b_2|^2)} } \no
&&~~~~~~~~~+{ {t({\bar b}_1(a_2(-a_4^2+|t|^2{\bar t}a_3b_1)+a_4|b_2|^2-{\bar 
t}a_3b_1(-a_4+{\bar t}b_2+t{\bar b}_2)))} \over{ 
(a_1a_3-|b_1|^2)(a_2a_4-|b_2|^2)} },\no
&&\bt \( { { {\bar b}_2\t a_2 -a_2\t {\bar b}_2} \over { a_2a_4-|b_2|^2} 
}\)={ {a_2(a_4({\bar t}b_1+t{\bar b}_1){\bar b}_2-a_3(|t|^2{\bar t}|b_1|^2 
+a_4{\bar b}_2)) } \over {(a_1a_3-|b_1|^2)(a_2a_4-|b_2|^2) } }\no
&&~~~~~~~~~~+{ {|t|^2a_1(a_2({\bar t}a_3^2 -a_4{\bar b}_2) +{\bar 
b}_2(-a_3^2+|b_2|^2))+{\bar b}_2(-({\bar t}b_1+t {\bar 
b}_1)|b_2|^2+a_3(|t|^2|b_1|^2+|b_2|^2)) }\over { 
(a_1a_3-|b_1|^2)(a_2a_4-|b_2|^2)} },\no
&&\bt \( { {a_2\t a_4 -{\bar b}_2 \t b_2} \over {-a_2a_4+|b_2|^2} }\)
={ {a_2(a_4^2({\bar t} b_1 +t {\bar b}_1)-a_3(a_4^2+|t|^4|b_1|^2)) }\over
{ (a_1a_3-|b_1|^2)(a_2a_4-|b_2|^2)} }\no
&&~~~~~~~~+{ {(-a_4({\bar t}b_1+t{\bar b}_1)b_2+a_3(|t|^2t |b_1|^2 
+a_4b_2)){\bar b}_2+|t|^2 a_1(a_2(|t|^2 
a_3^2-a_4^2)+(-t a_3^2+a_4b_2){\bar b}_2) }\over { 
(a_1a_3-|b_1|^2)(a_2a_4-|b_2|^2)} }
\ea
substantiated by the constraints coming from  
reality condition
\ba
&&(-{\bar t}a_1 +{\bar b}_1)(-t a_2 +b_2)+a_1(a_4-t{\bar b}_2)=1,\no
&&a_2(a_3-t{\bar b}_1)+(-ta_1+b_1)(-{\bar t}a_2+{\bar b}_2)=1,\no
&&a_2(a_3-{\bar t}b_1)+b_1{\bar b}_2=1,\no
&&{\bar b}_1b_2+a_1(a_4-{\bar t}b_2)=1,\no
&&a_2{\bar b}_1+a_1(-{\bar t}a_2+{\bar b}_2)=0,\no
&&(a_3-{\bar t}b_1)(-ta_2+b_2)+(b_1(a_4-t{\bar b}_2))=0,\no
&&(a_3-t{\bar b}_1)b_2+(-t a_1+b_1)(a_4-{\bar t}b_2)=0.
\ea
The examination of this constrained and coupled non-linear system is out of the scope of this paper, although it should be a integrable one from the zero curvature structure of $tt^*$ equation.

\section{ A sample residue calculation}
\setcounter{equation}{0}
 Here we
give a sample calculation  for the residue. \ba K=\la
(xy)^{n-1}|(xy)^{n-1}\ra=\int dx\wedge dy {(xy)^{2(n-2)} \over
{(x^{n-1}-ty)(y^{n-1}-tx)}} \nonumber \ea We evaluate the residue
iteratively, that is, the vanishing of the denominator gives
$y_{0i}=(tx)^{1/n-1}\omega^i$ with $\omega^{n-1}=1$ and
$y_1=x^{n-1}/t$. Then the residue integral of the $y$ becomes
 \ba
K&& =-\int dx \sum^{n-1}_{i=0} { {x^{2(n-2)}y_{0i}^{2(n-2)}} \over
{(n-1)y_{0i}^{n-2}(x^{n-1}-ty_{0i})} } + \int dx  { {
{x^{2(n-2)}y_1}^{2(n-2)}} \over {t (y_1^{n-1}-t x)}} \no
&&=-\sum^{n-2}_{i=0} {1\over {n-1}} \int dx { { x^{2(n-2)}
(tx)^{(n-2)/(n-1)}\omega^{i(n-2)}} \over
{x^{n-1}-t(tx)^{1/(n-1)}\omega^i}} +{1\over t} \int dx { {
(x^{n-1}/t)^{2(n-2)} x^{2(n-2)} } \over {(x^{n-1}/t)^{n-1}-tx}}
\nonumber \ea The first integral is zero since it can be brought
to the polynomial denominator  by change of variable $x=u^{n-1}$.
But then the $u$ contour does not close the poles when $x$ wrap a
circle.
The second integral becomes
\ba
&&(1/t) \int dx   { {t^{-2(n-2)}x^{2(n-2)(n-1+1)-1} } \over {
t^{1-n}x^{n(n-2)}-t}} \no &&=(1/t)t^{-2(n-2)}  \sum_{i=0}^{n(n-2)-1}   {{
x^{2(n-2)n-1} } \over t^{1-n} n(n-2) {x^{n(n-2)-1} }}\Big|_{x=x_i} \no &&=t^2,\nonumber
\ea
where $x_i$ is a solution of $x^{n(n-2)}=t^n$.

\vskip .5cm
\noindent {\bf \large Acknowledgement} \\
This work is  supported by KOSEF Grant R01-2004-000-10520-0.

\end{document}